\documentclass{emulateapj}
\usepackage{amssymb,amsmath,amsthm}
\usepackage{longtable}
\tabletypesize{\scriptsize}
\usepackage{color}
\usepackage{pdfcolmk}
\definecolor{DarkGreen}{rgb}{0.0, 0.5, 0.0}
\definecolor{purple}{rgb}{0.6, 0.0, 0.4}

\newcommand{\uv}{\mbox{$u$-$v$}}

\newcommand{\midka}{{31\,{\rm GHz}}}
\def\betaModel{$\beta$-model}

\newcommand{\rms}{{\it rms}}
\newcommand{\snr}{{\it snr}}

\def\planck{{\it Planck}}
\def\fgas{{$f_{gas}$}}

\newcommand{\be}{\begin{equation}}
\newcommand{\ee}{\end{equation}}

\def\plckA{{\it PLCKESZ G115.71+17.52}}
\def\plckB{{\it PLCKESZ G121.11+57.01}}
\def\plckC{{\it PLCKESZ G189.84-37.24}}

\def\mason09{M09}

\begin{document}
\title{CARMA follow-up of the northern unconfirmed \planck\ galaxy cluster candidates}

\author{
Stephen~Muchovej\altaffilmark{1,2},
Erik~Leitch\altaffilmark{1,3},
Thomas~Culverhouse\altaffilmark{1},
John~Carpenter\altaffilmark{2},
Jonathan~Sievers\altaffilmark{4}
}
\altaffiltext{1}{California Institute of Technology, Owens Valley Radio Observatory, Big Pine, CA 93513}
\altaffiltext{2}{California Institute of Technology, Department of Astronomy, Pasadena, CA 91125, USA}
\altaffiltext{3}{Department of Astronomy and Astrophysics, Kavli Institute for Cosmological Physics, University of Chicago, Chicago, IL 60637}
\altaffiltext{4}{CITA, University of Toronto, 60 St. George St., Toronto ON, M5S 3H8, Canada}

\begin{abstract}

We present CARMA observations of the three northern unconfirmed galaxy
clusters discovered by the \planck\ satellite.  We confirm the
existence of two massive clusters (\plckA\ and \plckB) at high
significance.  For these clusters, we present refined centroid
locations from the 31~GHz CARMA data, as well as mass estimates
obtained from a joint analysis of CARMA and \planck\ data.  We do not
detect the third candidate, \plckC, and place an upper limit on its mass
of ${M_{500} < \rm 3.2\times 10^{14}M_{\odot}}$ at 68\% confidence.
Considering our data and the characteristics of the \planck\ Early
Release SZ Catalog, we conclude that this object is likely to be a
cold-core object in the plane of our Galaxy.  As a result, we estimate
the purity of the ESZ Catalog to be greater than 99.5\%.
\end{abstract}
 
\keywords{galaxies: clusters: individual ({\it PLCKESZ G115.71+17.52}, {\it PLCKESZ G121.11+57.01}, {\it PLCKESZ G189.84-37.24}), techniques: interferometric  }

\section{Introduction}

Galaxy clusters are the most massive, gravitationally-bound structures
in the Universe.  Over a Hubble time, they form from the rare,
high-density peaks in the primordial density field on scales of
$\sim10~{\rm Mpc}$.  As the abundance of galaxy clusters depends
critically on the matter power spectrum and the expansion rate,
cluster surveys are a sensitive probe of cosmological parameters such
as the matter power spectrum normalization $\sigma_8$ and the dark
energy equation of state $w$.

The Sunyaev-Zel'dovich (SZ) effect is a spectral distortion of the cosmic
microwave background (CMB) radiation caused by inverse Compton
scattering of the CMB photons by electrons in the hot intra-cluster medium (ICM)
(\citealt{sunyaev70,sunyaev72}, see also
\citealt{birkinshaw1999}). The magnitude of the effect is proportional
to the integrated pressure of the ICM, i.e., the density of electrons
along the line of sight, weighted by the electron temperature. The integrated SZ
flux of a cluster is therefore a measure of its total thermal energy.

The change in the observed brightness of the CMB caused by the SZ
effect is given by
\begin{equation}
\label{y}
\frac{\Delta T_{\rm CMB}}{T_{\rm CMB}}=f(x)\int \sigma_{\rm T}n_e\frac{k_BT_e}{m_ec^2} dl \equiv f(x)y
\end{equation}
where $T_{\rm CMB}$ is the cosmic microwave background temperature (2.73 K), 
$\sigma_{\rm T}$ is the Thomson scattering cross section, $k_B$ is Boltzmann's constant, $c$ is the speed of light,  
and $m_e$, $n_e$, and $T_e$ are the electron mass, number density and temperature. 
 Equation~(\ref{y}) defines the Compton $y$-parameter.
The frequency dependence of the SZ effect is contained in the term
\begin{equation}
\label{f}
f(x)=\left(x\frac{e^x+1}{e^x-1}-4\right)\left(1+\delta_{\rm SZ}(x,T_e)\right) ,
\end{equation}
where $x\equiv h\nu/{k_BT_{\rm CMB}}$, $h$ is Planck's constant, and
$\delta_{\rm SZ}$ is a relativistic correction, for which we adopt the
\citet{itoh1998} calculation, valid to fifth order in
$k_BT_e/m_ec^2$. The SZ effect appears as a temperature decrement at
frequencies below $\approx 218~{\rm GHz}$, and an increment at higher
frequencies.

The redshift independence of the SZ effect in both brightness and
frequency (the ratio $\Delta T/T$ in equation~(\ref{y}) is independent
of the distance to the cluster) offers enormous potential for finding
high-redshift clusters.  Searches for massive galaxy clusters via the
SZ effect have the potential to produce cluster catalogs with a simple
mass selection, nearly independent of redshift if the angular
resolution of the observations is sufficient to resolve the cluster
\citep{carlstrom2002}.  As a result, several experiments have recently
conducted searches for galaxy clusters via their Sunyaev-Zel'dovich
(SZ) effect, e.g., the Sunyaev-Zel'dovich Array (SZA)
\citep{muchovej2011} -- now a part of the Combined Array for Research
in Millimeter-wave Astronomy (CARMA), the South Pole Telescope (SPT)
\citep{vanderlinde2010}, the Arcminute Microkelvin Imager (AMI)
\citep{amiSurvey}, and the Atacama Cosmology Telescope (ACT)
\citep{marriage2011}.  Most recently, the \planck\ space telescope has
begun to measure the CMB over the whole sky in 9 bands, and at lower
resolution ($\sim 5\arcmin$), to search for massive clusters of
galaxies via their SZ effect\citep{planckClusters}.

The \planck\ Early Release Compact Source Catalogue has identified
189 clusters of galaxies, including 20 previously unknown
clusters.  Of these, 11 have been confirmed using XMM Newton, and 1
was confirmed using a combination of AMI and WISE data
\citep{planckClusters}.  As a result, 8 objects from the catalog were
unconfirmed at the time of the \planck\ early release, and over the
past year various groups in the astronomical community have
sought to confirm their existence and infer properties about these
newly discovered objects.  In particular, the SPT was used to confirm
all cluster candidates in the southern sky and AMI targetted the two
northern-most clusters, confirming one of them in conjunction with
WISE \citep{story2011,hurley2011,wen2009}.  In this paper, we present
SZ follow-up observations obtained with CARMA of the three
clusters visible from the northern sky: \plckA, \plckB, and \plckC.

Whereas the \planck\ data are sensitive to the bulk SZ signal
(resolution of $\sim 5\arcmin$), measuring the pressure profile of
these clusters requires SZ follow-up with higher-resolution
instruments. As we demonstrate in this work, the combination of the
two data sets yields an improved estimate of the cluster mass, which is
of particular interest to the calibration of SZ observables to
intrinsic cluster parameters.  This paper is organized as follows: we
present a description of the data and the resulting maps in \S 2, and
derived cluster properties in \S 3 and \S 4.  We present a discussion
and conclusion in \S5 and \S 6, respectively.

\section{CARMA observations}
\label{sec:data}

\subsection{Observations and Reduction}

The data presented in this paper were collected in ten separate observations 
with the compact 31~GHz sub-array of the Combined Array for Research
in Millimeter-wave Astronomy (CARMA).  This compact sub-array,
formerly known as the Sunyaev-Zel'dovich Array (SZA), consists of
eight 3.5\,m telescopes operating from 27-35~GHz, arranged such that six
of the telescopes are in a compact configuration with two outlying
telescopes to allow identification and removal of compact sources.  Data from the six-element
compact array are referred to as {\it short-baseline} data below,
while the data from the two outlying telescopes are referred to as
{\it long-baseline} data.  The array layout is similar to that
presented in \cite{muchovej2007}, with the main difference being that
one of the long E-W baselines has been changed to a N-S baseline.

Over the time period from June 2011 to August 2011, each cluster was
observed for 4-5 hours about transit, in an array configuration
designed to minimize shadwoing by other antennas in the array
principally for sources at low declinations.  We require that clusters
are observed at an elevation greater than 30 degrees (to minimize
atmospheric contamination) for at least two hours during the day.
This limited our observations to the three unconfirmed \planck\
clusters in the northern hemisphere.  Cluster observations were
interleaved with observations of a strong unresolved source every 15
minutes to monitor variations in the instrumental gain.  \plckA\ was
observed over 4 tracks for a total of 9.3 hours of un-flagged
on-source data.  Likewise, we obtained 8.0 hours of unflagged
on-source data over 3 tracks on \plckB, and 5.8 hours of unflagged
data on \plckC\ obtained over 3 tracks.  Data were converted from the
MIRIAD format to Matlab, and calibrated in the same pipeline outlined
in \cite{muchovej2007}.  Absolute calibration is derived from
observations of Mars, using fluxes predicted by the most
up-to-date Rudy model scaled by 2\% to match the latest WMAP
measurements \citep{rudy1987, weiland2011}.  We estimate the flux
calibration to be good to 5\% via long-time monitoring of flux
calibrators used by the SZA.  In Table \ref{tab:obsTable}, we give
the pointing center of the cluster along with details of the
observations, including the synthesized beam sizes for both the short
and long baseline data.  We also present the achieved \rms \ flux
sensitivities for maps made with short and long-baseline data.
The effect of the array being in an orientation optimized for
low-declination sources is evident upon inspection of the
sensitivities achieved for each of the fields.  In particular, a
greater number of inner-array antennas are shadowed when observing
sources at higher declination.  As a result, observations of sources
at high declination can require a longer integration time to achieve
the same rms sensitivity as observations of low-declination sources.

\begin{deluxetable*}{lrrcccccccc}
\tabletypesize{\scriptsize}
\tablecolumns{8}
\setlength{\tabcolsep}{1mm}
\tablecaption{Cluster Observations}
\tablehead{
\colhead{Cluster Name}& \multicolumn{2}{c}
{\underline{Pointing Center (J2000)}}& \colhead{$\rm{t_{int}}$\tablenotemark{a}}& 
\multicolumn{2}{c}{\underline{Short Baselines (0-2k$\lambda$)} }&
\multicolumn{2}{c}{\underline{Long Baselines (2-8k$\lambda$)}}\\
\colhead{} & \colhead{$\alpha$}& \colhead{$\delta$} & \colhead{(hrs)} &
\colhead{beam($\arcsec\times\arcsec \angle$)\tablenotemark{b}}& 
\colhead{$\sigma$(mJy)\tablenotemark{c}} & \colhead{beam($\arcsec\times\arcsec \angle$)\tablenotemark{b}} & 
\colhead{$\sigma$(mJy)\tablenotemark{c}}
}
\startdata
\noindent \plckA\  & 22$^h$26$^m$24$^s$.89 &78$^{\circ}$18$^{\prime}$16.11$^{\prime\prime}$ & 9.3 & 118.2$\times$146.5 -34.5 & 0.41 & 12.7$\times$19.9 39.7 & 0.41\\
\noindent \plckB\  & 12$^h$59$^m$23$^s$.77 &60$^{\circ}$05$^{\prime}$24$.64^{\prime\prime}$ & 8.0 & 138.8$\times$146.0 -52.8 & 0.47 & 15.9$\times$19.8 43.3 & 0.50\\
\noindent \plckC\  & 03$^h$59$^m$45$^s$.80 &00$^{\circ}$06$^{\prime}$41.75$^{\prime\prime}$ & 5.8 & 105.2$\times$112.7 36.9 & 0.43 & 15.7$\times$23.2 37.1 & 0.51\\
\enddata

\label{tab:obsTable}

\tablenotetext{a}{On source integration time, unflagged data}
\tablenotetext{b}{Synthesized beam FWHM and position angle measured from North through East}
\tablenotetext{c}{Achieved \rms \ noise in corresponding maps}
\end{deluxetable*}

\subsection{Resulting Maps}

In the limit where sky curvature is negligible over the instrument's
field of view, the response of an interferometer on a single baseline,
known as a {\it visibility}, can be approximated by:
\begin{eqnarray}
\label{vis2}
V(u,v) = \int\!\!\int_{-\infty}^{+\infty}\!\!&&A_N(l,m)I(l,m) \\\nonumber
&\times& \exp \{ -2\pi j [ul + vm]\}{dl\,dm}, 
\end{eqnarray}
\normalsize where $u$ and $v$ are
the baseline lengths projected onto the sky, $l$ and $m$ are direction
cosines measured with respect to the $(u,v)$ axes,  $A_N(l,m)$ is the
normalized antenna beam pattern, and $I(l,m)$ is the sky intensity
distribution.  

As implied by equation~(\ref{vis2}), an image of the source intensity
multiplied by the antenna beam pattern, also known as a {\it dirty
map}, can be recovered by Fourier transform of the visibility data.
Note that in addition to modulation by the primary beam, structure in
the dirty map is convolved with a function that reflects the
incomplete Fourier-space sampling of a given observation.  This filter
function is the {\it synthesized beam}, equivalent to the point-spread
function for the interferometer.  A $clean$ map is an image from which
the synthesized beam pattern has been deconvolved, and the source
model reconvolved with a Gaussian fit to the central lobe of the
synthesized beam.  

\begin{figure*}
\begin{center}
\includegraphics[width=6.0in]{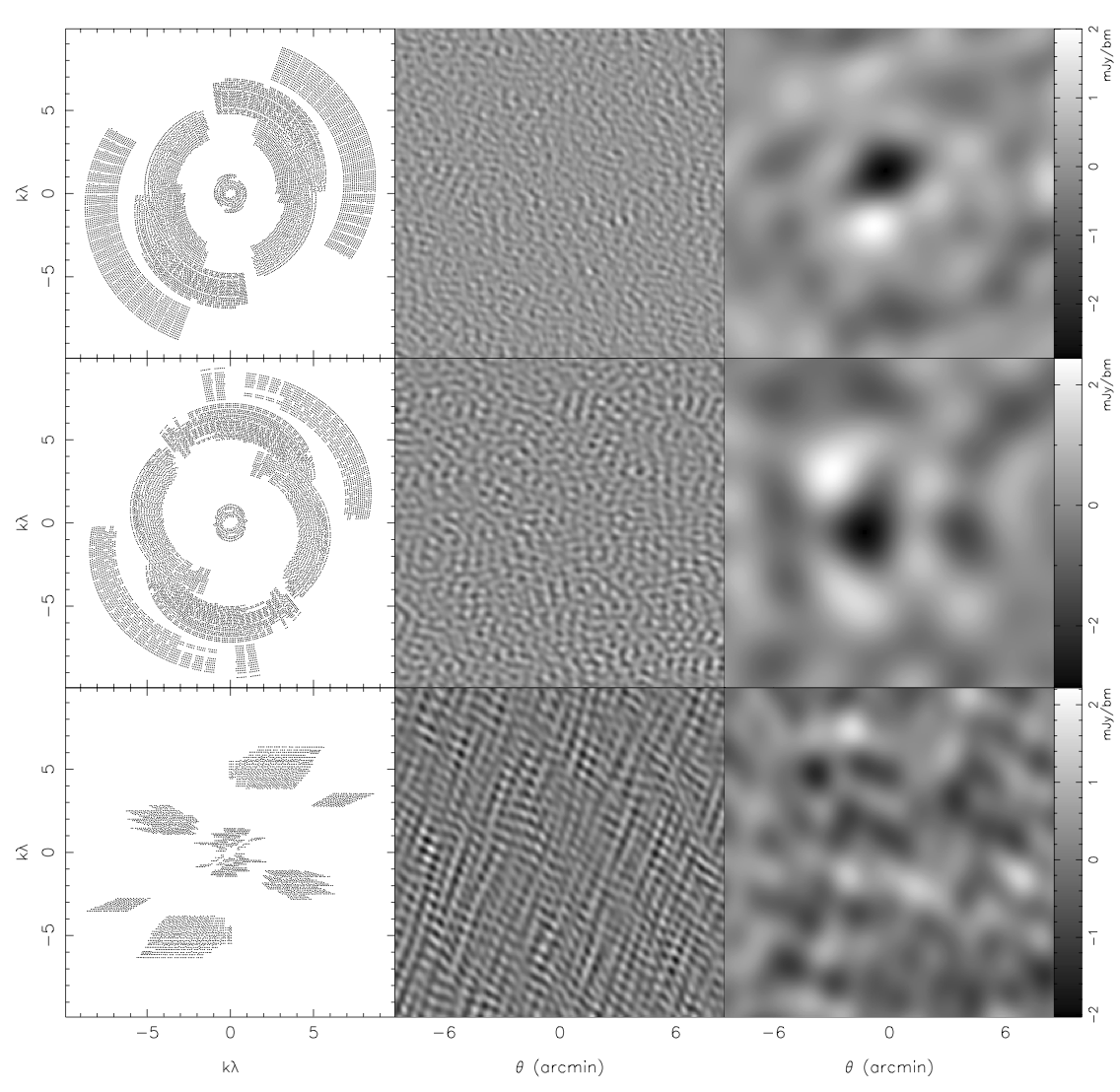}
\caption{ {\it Tow Row:} \uv-coverage, long baseline dirty map, and
  short baseline dirty map of data collected towards \plckA.  {\it
    Middle Row:} corresponding plots for field of \plckB.
  {\it Bottom Row}: same, but for \plckC.  Sensitivity and resolutions of observations are presented in Table \ref{tab:obsTable}.} 
\label{fig:plck12}
\end{center}
\end{figure*}

In the first column of Figure \ref{fig:plck12}, we present the
aggregate \uv\ coverage for observations of \plckA, \plckB, and
\plckC.  The second and third columns depict the corresponding {\it
dirty} maps obtained from the long and short baseline data,
respectively.  We identify two sources of emission in the field of
\plckA, corresponding to known sources from the NVSS catalog.
As no NVSS or FIRST coverage is available for the \plckB\ and
\plckC\ fields, we use the combination of the short and long-baseline
data to identify sources of emission directly from the SZA data at
greater than 3.5 times the map \rms\ level.  We identify one compact
source in the field of \plckB\ at four times the map \rms.  We do not
identify any sources of emission at a significance greater than 3.5
times the map \rms\ level in the field towards \plckC.  The
location and fluxes of these sources are presented in Table
\ref{tab:src_table}.  We note that the sparseness of our
Fourier sampling of \plckC\ does not hinder our ability to detect
sources of emission, as we are sensitive to scales as large as
4.5\arcmin.  The main effect of the sparse sampling is on the shape of
the {\it synthesized beam}, not our ability to detect extended sources
of emission.

As seen in the last column in Figure \ref{fig:plck12}, we detect an SZ decrement towards \plckA\ and \plckB\ at 6.1 and
6.0 times the \rms\ noise values in the map, respectively.  We detect no decrement
toward \plckC.  We
note that the images shown in Figure \ref{fig:plck12} are for display
purposes only, and that all source and cluster fluxes are fit directly
in the Fourier plane, as described in Section \ref{sec:param_est}.

\begin{deluxetable*}{lcrcrcccccl}

\tabletypesize{\scriptsize}
\tablecolumns{9}
\setlength{\tabcolsep}{0.9mm}
\tablecaption{Unresolved Radio Sources}
\tablehead{
\colhead{Cluster Field} &\colhead{$\#$} &\colhead{$RA$} & \colhead{$\sigma_{\rm RA}$} & \colhead{$DEC$}&\colhead{$\sigma_{\rm DEC}$} 
&\colhead{d\tablenotemark{a}}&\colhead{{\midka} Flux} & \colhead{1.4 GHz flux\tablenotemark{b}} & \colhead{$\alpha$} \\
\colhead{} & \colhead{}& \colhead{(J2000)} &\colhead{(s)} 
 & \colhead{(J2000)}& \colhead{(\arcsec)} & \colhead{(\arcmin)}& \colhead{(mJy)} & \colhead{(mJy)}& \colhead{(1.4/31 GHz)}}
\startdata
\plckA\  & 1 & $22^h26^m49^s.19$ &0.20& $+78^{\circ}16^{\prime}53^{\prime\prime}.8$ &3.1& 1.84 & $0.97\pm0.25$  & $30.88\pm 1.66$  & $1.11\pm 0.08$\\
        & 2 & $22^h26^m36^s.44$ &  --\tablenotemark{c} & $+78^{\circ}15^{\prime}25^{\prime\prime}.9$ & --\tablenotemark{c}& 2.90 & $0.47\pm0.21$  & $ 3.68\pm 0.55$  & $0.71\pm0.20$\\
\plckB\  & 1& $12^h59^m46^s.06$ & 0.27& $+60^{\circ}07^{\prime}09^{\prime\prime}.8$ &3.5& 3.28 & $1.76\pm0.43$ 
\enddata
\label{tab:src_table}
\tablenotetext{a}{Distance from observation pointing center}
\tablenotetext{b}{Integrated NVSS flux at 1.4 GHz}
\tablenotetext{c}{Due to low \snr, location fixed to NVSS centroid}

\end{deluxetable*}

\section{Cluster Parameter Estimation}
\label{sec:param_est}

All quantitative results presented in this paper are derived from
simultaneously fit models of the SZ cluster decrement and
contaminating sources, as detailed below.  In all cases, the model is
constructed in the image plane, multiplied by the primary beam, and
Fourier transformed, as indicated in equation~(\ref{vis2}).  The
resulting model visibilities are compared directly to the calibrated
visibility data.  In this way all fitting is done in the
Fourier-plane, where the visibility noise covariance is diagonal and
the spatial filtering of the interferometer is trivial to implement;
maps are used only for examination of the data and to identify cases
where contaminating sources are present.

The frequency-dependent shape of the primary beam used in the analysis
is calculated from the Fourier transform of the aperture illumination
of the telescopes, modeled as a Gaussian taper with a central
obscuration corresponding to the secondary mirror. The validity of
this model has been confirmed by holographic measurements.

We fit unresolved radio sources, hereafter referred to as a {\it point
sources}, as delta functions, parameterized by the
intensity at the band center, $I_\midka$, and a spectral index
$\alpha$ over our sixteen $500~\rm{MHz}$-wide correlator bands. The
point source intensity at frequency $\nu$ is then:
\begin{equation}
\label{int_ps}
I_{ps}(l,m) = I_\midka \left(\frac{\nu}{\midka}\right)^{-\alpha} \delta(l -
l^{\prime})\,\delta(m - m^{\prime}),
\end{equation}
where $l^\prime$ and $m^\prime$ are the coordinates of the point source on
the sky. From equations~(\ref{vis2}) and (\ref{int_ps}), it can be seen
that the visibility amplitude due to a point source is simply its
intensity, weighted by the normalized primary beam response at the
source location.

We model the cluster gas density by a spherical, isothermal
\betaModel, described by
\begin{equation} 
n_e(r) = {n_{e}}_0 \left ( 1 + {r^2 \over r_c^2} \right)^{-3 \beta/2},
\label{elec_dens}
\end{equation} where the core radius $r_c$ and the power law index
$\beta$ are shape parameters, and ${n_{e}}_0$ is the central electron
number density.  The model is a simple parameterization of the gas
density profile traditionally used in fitting X-ray
\citep[cf.][]{mohr1999a} and SZ data.  Although more complex
parameterizations can be shown to better reproduce fine details of the
density and temperature profiles of simulated clusters, when applied
to realistic data with the resolution of the SZA in this
configuration, the differences are irrelevant.  As a result, gas-mass
and total-mass estimates derived from the isothermal $\beta$-model
diverge from results obtained with more sophisticated pressure
profiles only at the cluster outskirts, and have been demonstrated to be consistent with each
other intermediate cluster radii (see Table 5 in \cite{mroczkowski2009}).

The corresponding SZ temperature decrement is given by
\begin{equation} {\Delta T}\left ( \theta \right ) = \Delta T_0 \left( 1 +  {\theta^2 \over
\theta_c^2} \right)^{{1\over 2} -{3\beta \over 2}},\label{eq:DT/T}
\end{equation} 
where $\theta = r/D_A $, $\theta_c = r_c / D_A$, and $D_A$ is the angular
diameter distance.
Under the assumption that the gas is isothermal, the temperature decrement at zero projected radius, $\Delta T_0$, is related to
${n_e}_0$ by

\begin{equation}
\label{neo}
{n_e}_0 = \frac{\Delta T_0}{T_{\it CMB}} \frac{m_e c^2}{f(x)k_B\sigma_T}\frac{1}{T_e}\frac{1}{\sqrt{\pi} D_A \theta_c }\frac{\Gamma(\frac{3\beta}{2})}
{\Gamma(\frac{3\beta}{2}-\frac{1} {2})}.
\end{equation}

Best-fit values for the model parameters are determined using a Monte
Carlo Markov Chain analysis \citep[][and references
therein]{bonamente2004,bonamente2006,laroque2006}.  The Markov chains
are a sampling of the multi-dimensional likelihood for the model
parameters, given the SZ data; the histogram of values in the chain
for each parameter is thus an estimate of the probability distribution
for that parameter, marginalized over the other model parameters.
The parameter $\beta$ was fixed to 0.86, consistent with the
average shape of massive clusters determined from the analysis of 15
massive clusters with the SPT \citep{plagge2010}.  This represents a
shift from previous joint analyses of X-ray and SZ observations
which traditionally used $\beta$ values of $2/3$ \citep[e.g.,][]{mohr1999a,
laroque2006}.

In Table \ref{tab:cluster_locs}, we present offsets from the \planck\
centroids determined for \plckA\ and \plckB.  For these and all other
quantities determined from the Markov chains, we quote the
maximum-likelihood value, with an uncertainty obtained by integrating
the distribution for that quantity to a fixed probability density,
until 68\% of the probability is enclosed.

\begin{deluxetable}{lcc}
\tabletypesize{\footnotesize}
\tablecolumns{3}
\renewcommand{\arraystretch}{1.1}
\tablecaption{CARMA Centroid Offsets from \planck\ }
\tablehead{
\colhead{Cluster Name} & \colhead{${\Delta{\rm RA}}$ (\arcsec)} & \colhead{${\Delta{\rm DEC}}$ (\arcsec)} }
\startdata
\plckA\    & $22.3^{+6.3}_{-12.7}$ & $70.4^{+11.1}_{-6.9}$ \\
\plckB\    & $84.5^{+18.0}_{-11.0}$ & $-15.1^{+10.1}_{-15.1}$\\
\enddata
\label{tab:cluster_locs}
\end{deluxetable}

\section{SZ Temperature and Mass Estimates}

\label{sec:mass_est}

In this section, we describe how the cluster electron temperature, gas
mass and total mass are determined from the Markov chains of model
parameters described in \S\ref{sec:param_est}.  

An estimate of the gas mass in the cluster can be obtained by
multiplying equation~(\ref{elec_dens}) by $\mu_e m_p$, the mean mass per electron of the ions in the plasma, and integrating the result to the desired radius:
\begin{equation}
\label{eq:gasMassInt}
M_{gas}(R)  = {\mu_e m_p  {n_{e}}_0}\int_{0}^{R}\left ( 1 + {r^2 \over r_c^2} \right)^{-3 \beta/2}  4\pi r^2 dr.
\end{equation}
The central electron density ${n_e}_0$ is a function of the electron
temperature $T_e$ (assumed to be constant) and the model parameters $\Delta T_0$, $\beta$ and
$\theta_{c}$, as given by equation~(\ref{neo}).

The total mass of the cluster can be estimated
by assuming hydrostatic equilibrium (hereafter HSE) and only thermal pressure support (i.e, no turbulent or rotational support).  For the electron
distribution given by equation~(\ref{elec_dens}), this approximation
yields an analytic solution for the total cluster mass contained
within a radius $R$ of:
\begin{equation}
\label{eq:hse}
M_{total}(R) = \frac{3k_BT_e\beta}{G\mu m_p} \frac{R^3}{r_c^2+R^2},
\end{equation}
where $G$ is the gravitational constant, $\mu m_p$ is the mean
molecular mass of the gas, and $r_c$ is the core radius, related to
$\theta_c$ by the angular diameter distance.  We adopt a value of
0.3\,$Z_{\odot}$ for the cluster metallicity when calculating both
$\mu_e$ and $\mu$, and assume a {$\Lambda$CDM} cosmology with
parameters fixed to those from the {\it WMAP} 7-year analysis in all
subsequent calculations \citep{larson2010}.

From equations (\ref{neo})--(\ref{eq:hse}), we see that if we assume a
value for the ratio of the gas mass to the total cluster mass,
hereafter referred to as the {\it gas-mass fraction}, $f_{gas}$, an
estimate of electron temperature can be inferred, allowing the masses
to be determined without reference to an {\it a priori} value for
$T_e$ \citep[c.f.,][]{joy2001, laroque2003}. We employ this method below to obtain cluster properties from
the SZ data.  For comparison, spectroscopically determined electron
temperatures from X-ray measurements can be used to estimate the gas
masses, total masses, and $f_{gas}$ directly from the Markov chains.
\label{sec:elec_temp}
A previous study of a sample of 38 massive
clusters obtained a mean of ${f_{gas} = 0.116\pm 0.005}$, from
masses evaluated within a radius of $R_{2500}$
\citep{laroque2006}. In the calculation of the gas temperature for a
single cluster, we therefore adopt a Gaussian distribution of \fgas\
with a mean of 0.116 and standard deviation of 0.030, where we have
scaled the reported error in the mean by $\sqrt{37}$ to approximate the
measured distribution of gas-mass fractions.

\begin{deluxetable*}{lccl|ll|ll}
\tabletypesize{\scriptsize}
\tablecolumns{7}
\tablecaption{Cluster Masses and ICM Properties Derived from SZ data}
\tablehead{
\colhead{} & \colhead{} & \colhead{} & \multicolumn{2}{c}{\underline{Quantities within $R_{2500(z)}$ }} & \multicolumn{2}{c}{\underline{Quantities within $R_{500(z)}$}}\\
\colhead{Cluster Name}& \colhead{Prior} & \colhead{$T_e$} & \colhead{$M_{gas}$} & \colhead{$M_{total}$} & \colhead{$M_{gas}$} & \colhead{$M_{total}$} \\
\colhead{}& \colhead{} & \colhead{(keV)} & \colhead{($10^{12}M_{\odot}$)} & \colhead{($10^{13}M_{\odot}$)} & \colhead{($10^{12}M_{\odot}$)} & \colhead{($10^{13}M_{\odot}$)}  
}
\startdata
\plckA\  & {\rm none}  & $5.3^{+1.1+0.2}_{-0.9-0.2}$  &  $29.2^{+4.6+2.3}_{-4.6-2.6}$  & $24.0^{+6.7+1.1}_{-5.4-3.2}$ &    $51.9^{+12.0+1.5}_{-9.0-1.2}$  & $53.7^{+18.0+5.9}_{-12.0-6.2}$  \\
\plckA\  & {\it Planck\tablenotemark{a}}  & $4.9^{+1.0+0.2}_{-0.6-0.2}$  &  $28.8^{+4.9+2.3}_{-4.9-2.6}$  & $21.6^{+8.1+1.1}_{-3.4-3.2}$ &   $51.8^{+11.2+1.5}_{-11.2-1.2}$  & $51.5^{+15.8+5.9}_{-10.5-6.2}$  \\
\plckB\  & {\rm none}  & $6.4^{+1.5+0.2}_{-1.5-0.2}$  &  $24.6^{+5.8+2.3}_{-8.2-0.9}$  & $15.9^{+10.5+3.2}_{-3.9-2.5}$ &   $81.0^{+15.0+2.1}_{-20.6-2.1}$  & $61.5^{+26.6+9.4}_{-11.8-7.2}$  \\
\plckB\  & {\it Planck\tablenotemark{b}}  & $5.7^{+1.1+0.2}_{-1.1-0.2}$  &  $24.8^{+5.4+2.3}_{-7.7-0.9}$  & $18.5^{+5.7+3.2}_{-6.6-2.5}$ &   $74.4^{+19.3+2.1}_{-19.3-2.1}$  & $58.1^{+18.0+9.4}_{-11.9-7.2}$  \\
\enddata
\label{tab:phys_params}
\tablenotetext{a}{$\theta$ = 14.47\arcmin, $\sigma_\theta$ = 7.333 from ESZ catalog}
\tablenotetext{b}{$\theta$ = 17.99\arcmin, $\sigma_\theta$ = 5.902 from ESZ catalog}
\end{deluxetable*}

Calculating the gas mass by integrating equation~(\ref{eq:gasMassInt})
requires knowledge of the redshift of the cluster (to determine the
physical radius over which to integrate).  
As no redshift information is available for these
objects, we marginalize over the redshift distribution of the
newly-discovered \planck\ clusters.  This distribution consists of 17 objects
whose redshifts are determined via either X-ray or optical follow-up,
with a median redshift value of 0.32 \citep{planckXmm, story2011,
planckClusters}.

We calculate the masses from the Markov chains by sampling the
distributions of ${\theta_c}$, $\Delta T_0$, ${f_{gas}}$ and
{$z$}, and solving for $T_e$ at an overdensity radius of $R_{2500}$ (where the estimates of ${f_{gas}}$ are determined).  The
resulting best-estimates for the electron temperature are presented in
Table \ref{tab:phys_params}.  Equipped with estimates of the electron
temperature, we can readily obtain estimates of the gas-mass and total
cluster mass from equations~(\ref{eq:gasMassInt}) and (\ref{eq:hse}).
In Table \ref{tab:phys_params} we present these values for \plckA\ and
\plckB, integrated to an overdensity radius of $R_{2500}$, and to
$R_{500}$, assuming that $f_{gas}$ is constant with radius.  The
overdensity radius $R_{\Delta}$ is defined as the radius at which the
mean density of the cluster is related to the critical density of the
Universe by a fixed density contrast $\Delta(z)$, where the density
contrast is assumed to scale with redshift like the mean
density of a virialized system, as determined from numerical
simulations \citep{bryan1998}.

An interferometer has no ability to constrain the size of an object
larger than the spatial scale of its shortest baselines.  For CARMA at
31~GHz, the instrument is insensitive to scales $\gtrsim10$\arcmin\ on
the sky.  The \planck\ satellite, on the other hand, cannot constrain
cluster models more compact than its highest resolution element
(namely 5~\arcmin), but can readily constrain the size of larger
objects.  As a result, we obtain the tightest constraints on the
cluster temperature and masses by including prior information on the
angular size of these clusters from the \planck\ satellite.  The
\planck\ ESZ catalog presents a angular extent from these clusters (at
$5\theta_{500}$) with an associated uncertainty.  The resulting masses
and temperatures, when this prior is included in the Markov chains,
are also shown in Table \ref{tab:phys_params}.  We see that
including the \planck\ prior reduces the 
statistical uncertainty in our determinations of gas temperatures by
25\%, and our final estimate of total masses by 15-30\%.

The choice of $\beta$ is one of the dominant systematic
uncertainties associated with our calculation.  This effect is more
pronounced on the cluster outskirts, where recent studies of the
average cluster profile have shown an increasing power law slope at
higher radii \citep{arnaud2010, sun2011}.  \cite{plagge2010}
determined a mean value of $\beta$ of ${\rm 0.86\pm0.09}$ from the
stacking analysis of 15 clusters.  To estimate the error
introduced by our choice of $\beta$, we repeat our analysis using
values of 0.77 and 0.95.  We see from Table
\ref{tab:phys_params} that this effect is largely negligible at the
inner radii of clusters, and leads to a roughly 10\% uncertainty at
larger radii.  We note that this uncertainty is still much smaller
than the statistical uncertainty in our mass estimate.

\section{Discussion}

\begin{figure*}
\begin{center}
\includegraphics[width=6.5in]{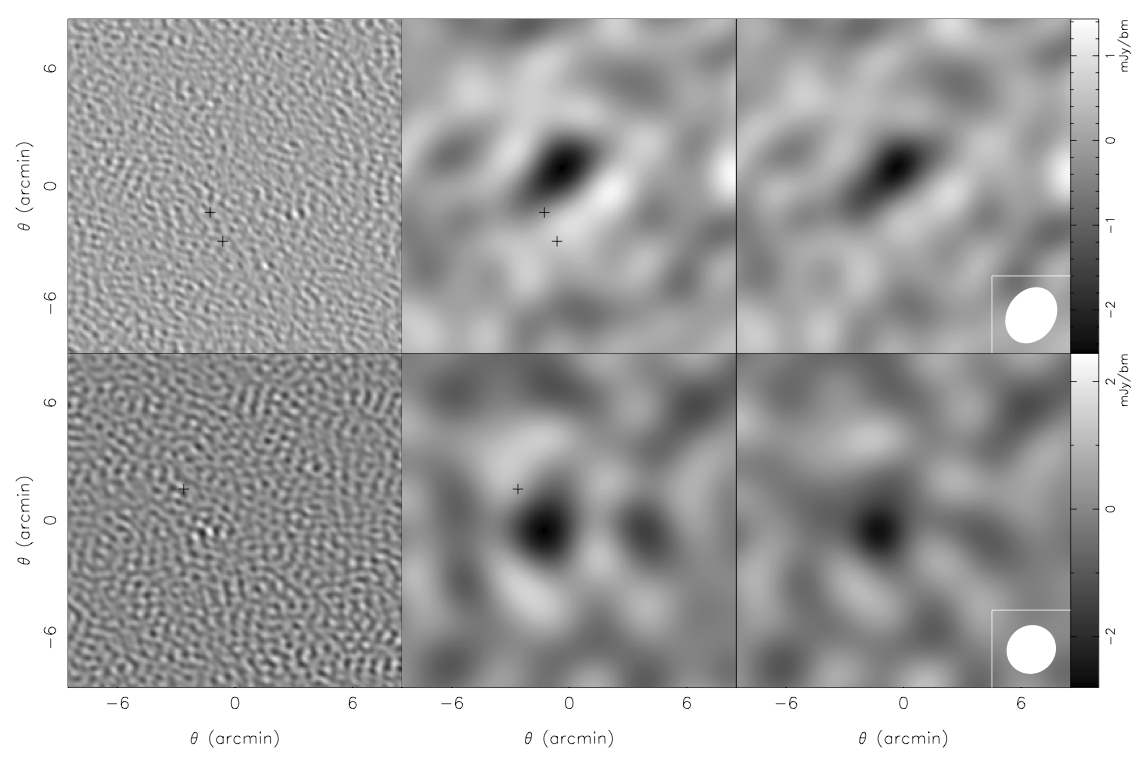}
\caption{{\it Top Row}: \plckA\ Long Baseline residual map once sources
of emission are removed from the data; Short Baseline residual map
once sources are removed; Cleaned map of \plckA. {\it Bottom Row}:
Corresponding images for \plckB.  Locations of sources removed from
the data are depicted by crosses.}
\label{fig:plck_clean}
\end{center}
\end{figure*}

\subsection{\plckA}

We confirm the presence of a massive galaxy cluster corresponding to
\plckA.  We determine the centroid of this cluster to be offset from
the \planck\ location by slightly more than an arcminute, at
RA~22:26:31.3, DEC~+78:19:28.7.  In the first two columns of Figure
\ref{fig:plck_clean} we present the long and short baseline {\it dirty
maps} of this cluster once sources of emission are removed.  In the
last column of the first row, we present the resulting cleaned image
of this cluster.  We estimate the mass of this cluster to be
${\rm M_{500} = 5.2 ^{+1.6+0.6}_{-1.1-0.6} \times 10^{14}
M_{\odot}}$, where the first set of errors correspond to the 1$\sigma$
statistical errors and the second set to the systematic uncertainty
due to our choice of $\beta$ (presented in Table
\ref{tab:phys_params}).  This value is consistent with the median mass
of clusters released in the \planck\ ESZ catalog
\citep{planckClusters}.  We note that the inclusion of the \planck\
prior in our analysis improves our mass estimate by 15\%, comparable
to the error associated with our choice of $\beta$.  As discussed
in \S\ref{sec:mass_est}, the mass estimate was obtained assuming
the redshift distribution of the newly-discoverd
\planck\ clusters.  As the SZ observations provide no information on
the redshift of the cluster, we present our determination of the mass
of this cluster as a function of redshift in Figure \ref{fig:plck_z}.
We note that our final mass estimate for this cluster is consistent
with that of the median redshift of the newly discovered \planck\
clusters, namely 0.32.

We note that this field has also been observed with the Arcminute
Microkelvin Imager (AMI), however in the presence of overwhelming
source contamination at 15~GHz, AMI was unable to to detect an SZ
decrement \citep{hurley2011} and confirm this cluster.
The CARMA data thus provide the first confirmation of this newly
discovered cluster.

\subsection{\plckB}

\begin{figure*}
\begin{center}
\begin{tabular}{cc}
\includegraphics[width=3.5in]{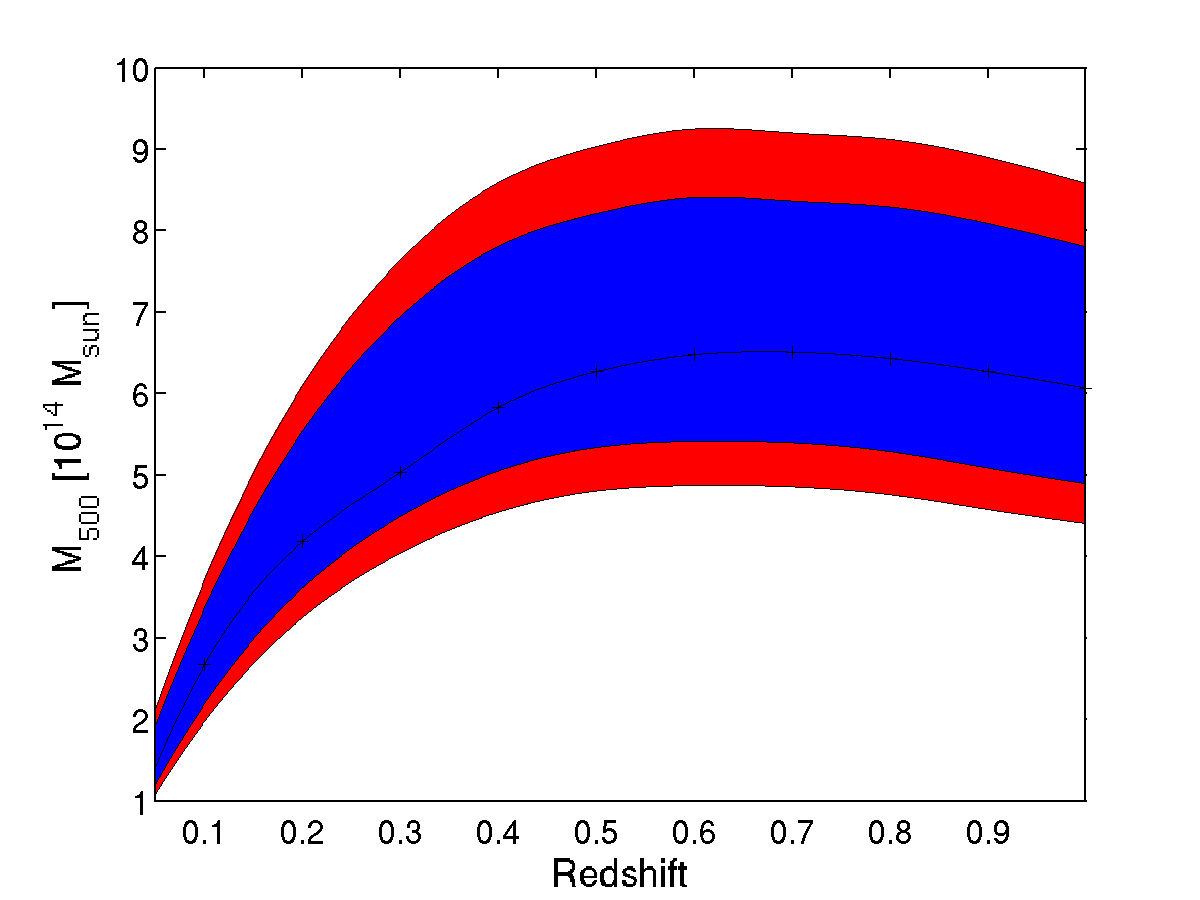} & \includegraphics[width=3.5in]{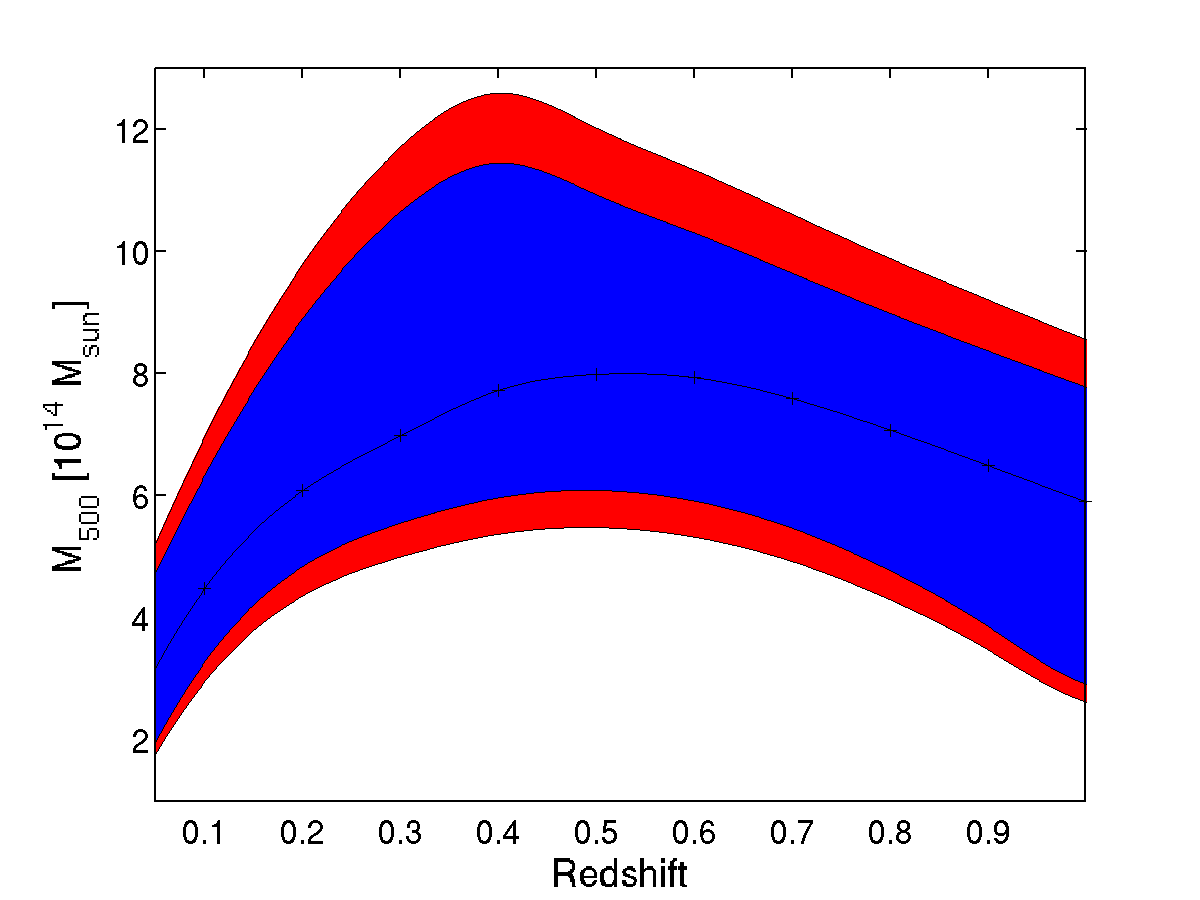} 
\end{tabular}
\caption{{\it Left}: Mass estimate of \plckA\ as a function of
redshift.  The blue shaded region indicate the 1$\sigma$ errors on the
most likely value of the mass (center line), and the red shaded region
is an estimate of the error due to the choice of $\beta$.  {\it
Right}: Same plot, but for \plckB.  We note that our final mass
estimates are consistent with the clusters being at a redshift
$\sim$0.32, the median redshift value of the newly-discovered \planck\
clusters.}
\label{fig:plck_z}
\end{center}
\end{figure*}

We detect a significant SZ decrement toward \plckB, confirming its
existence as a massive cluster.  We estimate the mass of this cluster
to be ${\rm M_{500} = 5.8 ^{+1.8+0.9}_{-1.2-0.7}\times 10^{14} M_{\odot}}$, and find
its centroid to be at RA~12:59:35.8, DEC~+60:05:09.1.  The inclusion of the \planck\ prior on the angular extent of this cluster reduces the uncertainty on our mass estimate by $\sim$ 28\%.  The cleaned image of
this cluster, with a single source of emission removed from the field,
can be found in the last panel of the second row in Figure
\ref{fig:plck_clean}.  As no redshift information is available for
this cluster, in the right panel of Figure \ref{fig:plck_z} we
present the estimated mass of this cluster as a function of redshift.

This cluster was previously confirmed with a 107-hour observation with
AMI \citep{hurley2011}.  We note that the cluster is detected with
comparable significance in the 8-hour CARMA track, and that the
determination of the cluster centroid agrees with that determined from
AMI to 22\arcsec (by comparison, the quoted accuracy on the AMI centroid is 20\arcsec).

\subsection{\plckC}

We detect no SZ decrement at the location of \plckC.  Furthermore,
as can be seen in Figure \ref{fig:plck12}, the field is free of source
contamination.  A non-detection of a genuine cluster in a 6-hour track
with CARMA would require either a low-mass compact cluster (SZ signal weak), or an extended, low-redshift cluster
(SZ signal resolved out).

\begin{figure}
\begin{center}
\includegraphics[width=3.5in]{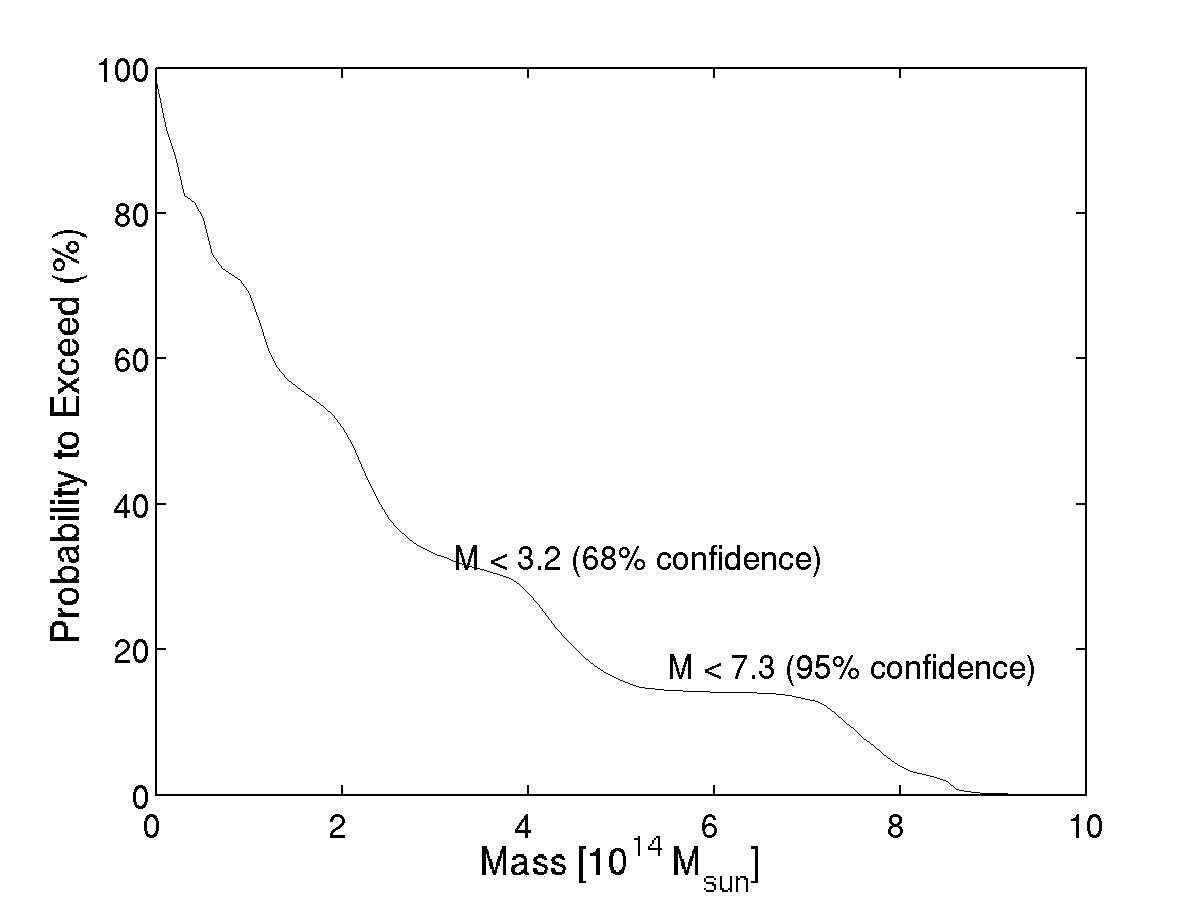}
\caption{Mass limit (${\rm M_{500}}$) on \plckC\  obtained assuming a
compact cluster within 1\arcmin\  of the location indicated by the \planck\ ESZ catalog}
\label{fig:plck3_lim}
\end{center}
\end{figure}

Under the assumption that a cluster is present within a 1 arcminute
radius of the \planck\ coordinate, and that it subtends the typical
scales of clusters,
we can place an upper
limit on the mass of the cluster, given our data.  A Markov chain is
run as described in Section \ref{sec:param_est}, and the formalism of
Section \ref{sec:mass_est} is applied to determine the distribution of
masses allowed by our data.  Under these assumptions, we can place an
upper limit on the cluster mass (${\rm M_{500}}$) of ${\rm 3.2{+0.3}\times
10^{14} M_{\odot}}$ at 68\% confidence, where the uncertainty is due to the choice of $\beta$, as seen in Figure
\ref{fig:plck3_lim}.

We note however that the \planck\ data indicate a size of 62.5\arcmin\
at ${\rm 5\theta_{500}}$.  An object this large would be undetectable
(resolved out) by the interferometer, so it is not surprising that the
CARMA data are consistent with noise, whatever the nature of the
source seen by \planck.  If this is a cluster, however, its angular
extent indicates that it is nearby (${ z \ll 0.1}$), and the $Y_{\rm
500}$ estimated from the \planck\ data implies an X-ray luminosity
several times larger than either \plckA\ or \plckB\
(\cite{wmapSzXrayScaling}, \cite{planckSzXrayScaling}), a source
easily detectable with ROSAT.  Yet the measured signal in RASS toward
this object, integrated over the \planck\ aperture, is consistent with
noise, and a factor of $3-6$ lower than toward the compact clusters
\plckA\ and \plckB\ \citep{planckClusters}.  The interpretation of
this source as a nearby cluster would therefore require unusual
conditions in the ICM to produce little or no central condensation,
leading to the selective suppression of X-rays relative to the SZ
signal.  Inspection of images from the Sloan Digital Sky Survey also
reveal no evidence for an over-abundance of galaxies consistent with
nearby clusters.

As a result, we believe that the most natural explanation for this
source is the contamination discussed in \cite{planckClusters}, where
it is noted that the prevalence of IR sources emitting above 217~GHz,
dust emission and cold cores was found to be higher than expected.
\planck\ identified many cool core objects near the Galactic plane,
including a southern region around Galactic longitude of 180 extending
south to longitude of $-45^\circ$ \citep{planckColdCore}, in which this
object lies.  The inclusion of data from the low-frequency instrument
(where the SZ signal, characterized by a decrement, can be readily
distinguished from a thermal spectrum) in the \planck\ cluster-finding
algorithm will clarify the nature of this source.

\section{Conclusion}

Of the new cluster candidates identified in the \planck\ Early
Release Compact Source Catalogue, three are visible in the northern
sky: \plckA, \plckB\ and \plckC.  From June-August 2011, we obtained
31~GHz observations of these candidates with the CARMA interferometer,
with a total of $5-10$ hours of observation per source.

SZ decrements are detected with high significance toward both \plckA\
and \plckB; we present refined centroid locations and mass estimates
at $R_{2500}$ and $R_{500}$ for each of these clusters.  Masses are
determined from the SZ data via an MCMC analysis, by assuming a
distribution for the gass-mass fraction from previous studies of
massive clusters, and by marginalizing over the redshift distribution
of the newly-discovered \planck\ clusters.  These masses
represent the first joint-analysis of Planck and interferometric SZ
data.  Masses were determined using the Planck priors on the size of
the clusters, resulting in mass uncertainties of roughly 20\%. An
extension of this work to a larger sample of clusters already observed
with CARMA will help tighten our constraints on SZ-scaling relations.
These data represent the first confirmation of \plckA, and the first
mass estimate for either cluster.

No SZ decrement was detected in the CARMA observations toward \plckC.
Given the non-detection, we can restrict the mass of a compact cluster
at this location to be less than $3.2\times 10^{14} M_{\odot}$ at 68\%
confidence.
However, the \planck\ data suggest that the
source is quite large, in which case it is not surprising that nothing
is seen in the CARMA data, which is insensitive to objects larger than $\sim10$\arcmin.  
Given its size, the object would have to
be nearby, which makes it unlikely that it would have escaped
detection in ROSAT if it is a genuine cluster.  We conclude that the
source is likely to be a dusty 'cold-core' object associated with the
Galactic plane.

The steep decline of the radio-source population with frequency makes
the intrinsic contribution of contaminating sources to the 31~GHz
CARMA data quite small \citep{muchovej2010}; a total of three compact
sources were removed from the observations of \plckA\ and \plckB.  The
hybrid array configuration allows these sources to be cleanly removed
from the short-baseline data with little impact on the final cluster
parameters.  In the case of \plckC\, there is no evidence for contaminating
sources present in the data.

This work, combined with follow-up with XMM-Newton
\citep{planckXmm}, a combination of AMI and WISE
\citep{planckClusters, hurley2011}, and SPT observations of
unconfirmed southern sources \citep{story2011}, confirms all
newly-discovered clusters in the \planck\ ESZ catalog, with the
exception of \plckC.  Under the assumption that this is not a
genuine cluster, we conclude that the purity of the ESZ catalog is
better than 99.5\%.

\acknowledgements We thank the staff of the Owens Valley Radio
Observatory and CARMA for their outstanding support.  We especially
would like to thank John Carlstrom for his efforts in spearheading the
construction and operation of the SZA, and for useful comments on the
manuscript.  Support for CARMA construction was derived from the
Gordon and Betty Moore Foundation, the Kenneth T. and Eileen L. Norris
Foundation, the James S. McDonnell Foundation, the Associates of the
California Institute of Technology, the University of Chicago, the
states of California, Illinois, and Maryland, and the National Science
Foundation. Ongoing CARMA development and operations are supported by
the National Science Foundation under a cooperative agreement (grant
AST 08-38260), and by the CARMA partner universities (in particular
NSF grant AST 0838187). SM gratefully acknowledges support from an NSF
Astronomy and Astrophysics Fellowship.

{\it Facilities:} SZA, CARMA

\bibliographystyle{apj}

\bibliography{ms_arxiv}

\end{document}